# Should bike sharing continue operating during the COVID-19 pandemic? Empirical findings from Nanjing, China

**Mingzhuang Hua [a, b, c], Xuewu Chen [a, b, c, *], Long Cheng [d], Jingxu Chen [a, b, c]**

[a] Jiangsu Key Laboratory of Urban ITS, Southeast University

[b] Jiangsu Province Collaborative Innovation Center of Modern Urban Traffic Technologies, Southeast University

[c] School of Transportation, Southeast University

[d] Geography Department, Ghent University

* Corresponding author: Dongnandaxue Road #2, 211189 Nanjing, China
Telephone: +86 025 5209 1280; E-mail: chenxuewu@seu.edu.cn

**Abstract**

Coronavirus disease 2019 (COVID-19) has triggered a worldwide outbreak of pandemic, and transportation services have played a key role in coronavirus transmission. Although not crowded in a confined space like a bus or a metro car, bike sharing users will be exposed to the bike surface and take the transmission risk. During the COVID-19 pandemic, how to meet user demand and avoid virus spreading has become an important issue for bike sharing. Based on the trip data of bike sharing in Nanjing, China, this study analyzes the travel demand and operation management before and after the pandemic outbreak from the perspective of stations, users, and bikes. Semi-logarithmic difference-in-differences model, visualization methods, and statistic indexes are applied to explore the transportation service and risk prevention of bike sharing during the pandemic. The results show that pandemic control strategies sharply reduced user demand, and commuting trips decreased more significantly. Some stations around health and religious places become more important. Men and older adults are more dependent on bike sharing systems. Besides, the trip decrease reduces user contact and increases idle bikes. And a new concept of *user distancing* is proposed to avoid transmission risk and activate idle bikes. This study evaluates the role of shared micro-mobility during the COVID-19 pandemic, and also inspires the blocking of viral transmission within the city.

**Keywords:** COVID-19; Bike sharing; Travel demand; Transmission risk

# 1. Introduction

Coronavirus disease 2019 (COVID-19) is an infectious disease caused by the Severe Acute Respiratory Syndrome coronavirus 2 (SARS-CoV-2) (Gorbalenya et al., 2020). COVID-19 outbreak was first discovered in Wuhan, China, in December 2019. The confirmed cases have been subsequently diagnosed all over the world. Because of the alarming levels of spread and severity, the World Health Organization has characterized COVID-19 outbreak as a pandemic (World Health Organization, 2020). As of December 4, 2020, COVID-19 pandemic has caused 65,220,557 confirmed cases and 1,506,157 deaths (Johns Hopkins University, 2020). COVID-19 symptoms may appear 1-14 days after exposure to the virus, including fever, cough, and breathlessness (China Daily, 2020). Transmission of COVID-19 can occur through droplets and touching the surface with the coronavirus. COVID-19 poses a huge challenge for the travel demand and service safety of urban transportation.

Different modes of urban transportation have experienced various changes and employed different measures during the COVID-19 pandemic. With many people crowded in confined spaces, the transmission risk through public transport is very high. There is a significant association between acute respiratory infection and bus or tram use before symptom onset (Troko et al., 2011). So public transport agencies require riders to wear masks (BBC News, 2020), adjust operating services (Brown, 2020), and even shut down in many cities (Hubei Daily, 2020). Public transport has experienced a sharp decline of trip amount, such as the 80% decrease of public transport travel in Budapest, Hungary (Bucsky, 2020). Besides, ride-sharing and taxi could be high-risk environments for virus transmission. In addition to mask-wearing requirement and exchanging critical information (World Economic Forum, 2020), plastic protective films are also installed in cars (China News, 2020).

Shared micro-mobility (SMM) is the latest trend of urban transportation, and bike sharing is the core component of SMM. The SMM refers to shared-use fleets of small, manually or electrically powered vehicles (National Association of City Transportation Officials, 2020). Examples include electric scooter sharing, station-based bike sharing (SBBS), dockless bike sharing (DBS), and electric bike sharing. The SMM plays a key role in urban transportation amidst this pandemic. Unlike public transport and taxi, bike sharing provides services in open spaces and causes fewer safety concerns. During the COVID-19 shutdown period, bike sharing in Wuhan became the most frequently used travel mode and provided healthcare staff essential mobility services (CNR News, 2020). However, some cities and companies still stopped SMM services after the outbreak because of the safety concerns about virus transmission. Lime and Bird suspended their services of bike sharing and electric scooter sharing in many cities worldwide (Plautz, 2020; Gauquelin, 2020). The SMM faces a tough trade-off between operational service and health safety during COVID-19.

COVID-19 has caused a global health crisis in the past months, and transportation service has been profoundly affected as a result. However, there are very few studies exploring the role of bike sharing during the pandemic. And pandemic control measures on urban transportation still lack research. Furthermore, the outbreak coincided with the 2020 Chinese New Year (CNY) holiday, making it difficult to assess the impact of the pandemic on bike sharing. This paper estimates the travel demand and transmission risk of bike sharing during COVID-19 with the trip data of Nanjing Public Bicycle. This paper aims to help promote the SMM operation and reduce virus transmission risk, by comprehensively analyzing bike sharing service from multiple perspectives of users, stations, and bikes. To the best of our knowledge, this paper is the first paper focusing on both user demand and transmission risk of bike sharing during the COVID-19 pandemic.

This paper is divided into seven sections. Following this introduction, the second section presents the literature review about COVID-19 and transportation. The third section describes the data sources and processing procedures. The fourth and fifth sections present the travel demand change based on station and user perspectives. The sixth section analyzes the COVID-19 transmission risk through shared bikes. In the end, the last section presents the conclusion and discussion.

## 2. Literature review

### 2.1. COVID-19 and transportation

COVID-19 is a respiratory disease, which is generally susceptible to the whole population but more dangerous for weaker people. During hospital admission, the main diagnosis of patients is pneumonia, followed by Acute Respiratory Distress Syndrome (ARDS) and shock (Guan et al., 2020). Using a 14-day-delay adjusted estimation, the mortality rate of COVID-19 is 5.7% (Baud et al., 2020). Compared with survivors, the deaths are more likely to have chronic medical illnesses and are older (Yang et al., 2020). Based on the integration of extensive travel data and reliable infection data, the basic reproductive number $R_0$ is adjusted to be 5.7 (Sanche et al., 2020). The communicable period of COVID-19 patients could be up to 21 days, and asymptomatic virus carriers have non-negligible transmission potential (Hu et al., 2020). SARS-CoV-2 could remain infectious in aerosol for hours and on the surfaces of stainless steel and plastic for up to 3 days (van Doremalen et al., 2020).

Pandemic control strategies have been applied to contain and mitigate the COVID-19 outbreak. These measures include quarantine of infected cases, self-isolation, social distancing (Chen et al., 2020), cleaning touched surfaces, and wearing personal protective equipment (PPE) such as masks and gloves. Highly effective contact tracing and case isolation is enough to control COVID-19 outbreak in most scenarios (Hellewell et al., 2020). The social distancing strategy could be effective to reduce transmission and mitigate pandemic (Fong et al., 2020).

Transportation services are the key factors that lead to the rapid spread of COVID-19, and restricting transportation strategy is useful in slowing the pandemic spreading. Zhang et al. (2020) find that frequencies of high-speed trains and air flight services out of Wuhan are significantly related to the amount of infected cases. Because of the strong relationship between aviation services and pandemic spreading, air travel data are used to estimate the true size of Italian outbreak (Tuite et al., 2020). Chinazzi et al. (2020) confirm that international travel restrictions slow virus spreading around the globe. Tian et al. (2020) find that suspending public transport is related to the reduction of COVID-19 cases.

There have been some exploratory studies about the impacts of COVID-19 on transportation, but urban transportation and shared mobility are rarely considered. Suau-Sanchez et al., (2020) discover that the impact of COVID-19 on international air transport has been stronger than that on domestic airlines. Loske (2020) finds that the increasing freight volume for dry products in retail logistics depends on the increase of infected cases. Mogaji (2020) recognizes that transportation in Lagos, Nigeria were affected by this pandemic, including increased travel cost, shortage of travel mode, and traffic congestion.

Some studies have discussed transportation service supply during COVID-19 from a qualitative perspective. Budd and Ison (2020) propose the concept of Responsible Transport for post-COVID management, and stress that the individuals should assess the risk of their trips and take action accordingly. Musselwhite et al. (2020) suggest that reducing hypermobility of the transportation network and focussing on local connectivity could address virus threat. De Vos (2020) recommends cycling can maintain satisfactory levels of health and wellbeing, and active travel should be encouraged to cope with the problems of social isolation and limited physical activity.

Existing research between COVID-19 and transportation still has many gaps, and several questions need to be answered. COVID-19 can be transmitted through aerosol and surface, so bike sharing in open spaces is much safer than public transport and taxi in confined spaces. Bike sharing increases its modal share among all travel modes and provides critical mobility services during the outbreak. There are very few studies focusing on SMM amidst the pandemic. And the transmission risk through these shared vehicles has not been evaluated quantitatively.

## 2.2. Bike sharing

Extensive studies have explored improving the planning and operation of bike sharing, to provide better services for users. Previous studies about bike sharing can be generally classified into three categories: (1) user demand; (2) bike rebalancing; (3) parking allocation.

User demand is the base of service improvement, and user demand studies are mainly done through questionnaire surveys or trip data. Some studies focus on user behavior of bike sharing, such as the factors affecting demand (El-Assi et al. 2017; Wang et al. 2018; Kaplan et al., 2018; Biehl et al., 2019; Ma et al. 2020). McKenzie (2019) compares electric scooter sharing and bike sharing in Washington, D.C., and found that bike sharing is primarily used for commuting but electric scooter sharing is not. Demand prediction is also a key part of bike sharing research. The traditional regression models (Faghih-Imani et al. 2017; Almannaa et al. 2020) and the novel machine learning methods (Xu et al. 2018; Li and Zheng 2020) are both used in predicting bike sharing demand.

Bike rebalancing is to ensure that users can have bikes for riding and spaces for parking, including truck rebalancing and user incentive rebalancing. Truck rebalancing has two types: dynamic truck rebalancing and static truck rebalancing. Static truck rebalancing is a static vehicle routing problem using trucks to rebalance bikes (Ho and Szeto 2017; Bruck et al. 2019; Huang et al. 2020). And dynamic truck rebalancing is a dynamic vehicle routing problem (Caggiani et al. 2018; Agussurja et al. 2019; Tian et al. 2020b). User incentive rebalancing also attracted wide attention, encouraging users to rebalance bikes with monetary incentives (Haider et al. 2018; Zhang et al. 2019a; Duan and Wu 2019).

In SBBS, parking allocation is station planning and dock allocation, which is an optimal programming problem (Lin et al. 2018; Loidl et al. 2019; Caggiani et al. 2019; Soriguera and Jiménez-Meroño 2020). In DBS, parking allocation is where and how many park spaces should be allocated. Some studies detect the parking hotpots of DBS with machine learning methods, but the amount of parking spaces in each place is not solved well (Liu et al. 2018; Zhang et al. 2019). Hua et al. (2020) apply clustering of DBS trips to identify virtual stations, then recognize the maximum bike number in a day as the suitable amount of parking spaces in each station, finding that a DBS bike requires about 1.2 parking spaces.

Most existing studies on bicycle sharing are about normal operation, and there are only very few papers focusing on bike sharing amidst the COVID-19 pandemic. Teixeira and Lopes (2020) find the possible evidence on a modal transfer from the subway to bike sharing in New York City during COVID-19. Nikiforiadis et al. (2020) conduct a questionnaire survey of 223 people in Thessaloniki, Greece, and the results show that COVID-19 will not affect the user amount of bike sharing. The user demand and operational service of bike sharing during the COVID-19 pandemic remains to be studied.

## 3. Data

### 3.1. Data sources

Nanjing is the capital of Jiangsu Province and one of the super-large cities in China. This city has a population of 8.5 million, and a city area of 782.3 square kilometers. Nanjing Public Bicycle (NPB) started the SBBS service in 2013. As of February 2020, NPB had 53,200 bikes and 1,500 stations in operation. Nanjing has three DBS companies including Mobike, Didibike, and Hellobike, with a total of 350,600 free-floating bikes. The visualization of the service supply of bike sharing is shown in Figure 1. The DBS distribution is based on the position of each DBS bike at midnight. The distribution of NPB stations and DBS bikes coincides, except for some locations in the suburban area.

In Nanjing, the first covid-19 case was found on January 23, 2020. The last case in Nanjing was found on February 18, 2020, and there has been no case since then. Nanjing's urban transportation suffered the biggest impact in February 2020, and the monthly trips compared with 2019 falling by 90.6% in metro, 93.0% in bus, and 90.3% in taxi. The trip decline is consistent with people's extreme concern about levels of hygiene on public transport (Beck and Hensher, 2020). After February, Nanjing's urban transportation gradually recovered. In October 2020, the monthly trips compared with 2019 falling by 12.4% in metro, 27.8% in bus, and 12.4% in taxi. For Nanjing, February 2020 is the most critical and most affected month during the COVID-19 pandemic.

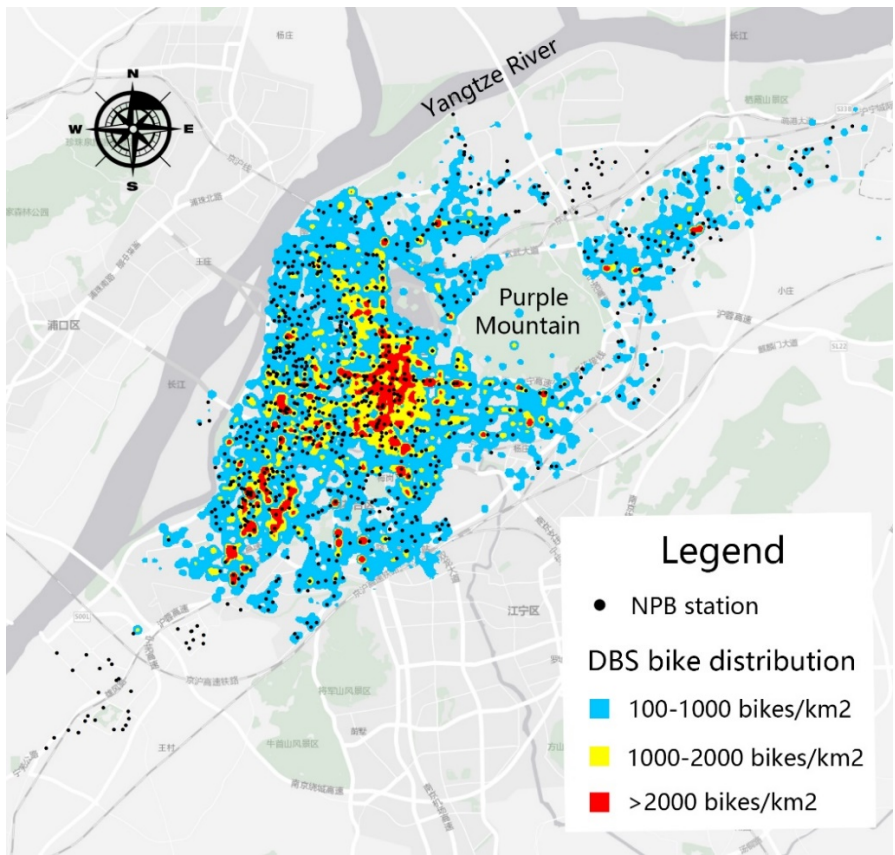

Figure 1. Spatial distribution of SBBS stations and DBS bikes in Nanjing.

Four years of SBBS trip data from March 2016 to February 2020 is provided by Nanjing Public Bicycle Company Limited. The raw SBBS dataset has 146.7 million trip records, with a file size of 19.9 GB. The fields of NPB trip data include: user ID, bike ID, departure time, arrival time, origin station, and destination station. The personal data of all 590,500 users during the four years were also provided, including user ID, gender, and age. The personal data are anonymized after data desensitization. Besides, the station list displays station name, latitude, and longitude. The latitude and longitude of NPB stations use the WGS-84 coordinate system.

Three kinds of defective SBBS trip data are eliminated. Data defects include (1) 2.5 million trip records that have no user ID, (2) 1.1 million trip records, of which the origin station is outside the station list, (3) 0.1 million trip records, of which the departure time or arrival time do not have the time precision to a second. After the data cleaning, 143.0 million SBBS trips are extracted for the analysis.

Four weeks of DBS trip data is provided by Nanjing Transportation Bureau. The two fortnights consist of two weeks in January 2020 and two weeks in February 2020. The raw DBS dataset has 4.8 million trip records, with a file size of 598 MB. The fields of DBS trip data include: bike ID, company ID, departure time, arrival time, origin longitude, origin latitude, destination longitude, and destination latitude. The personal information of DBS users was not provided.

Three kinds of defective DBS trip data are removed. (1) The departure time or arrival time of 954,000 records is missing. (2) The origins or destinations of 269,000 records are missing. (3) The origins or destinations of 2,000 records are not in Nanjing. Finally, 3.5 million DBS trips are used in the study.

Points of interest (POI) data were crawled from Auto Navi Map to explore the relationship between land use and bike sharing. Auto Navi Map is one of the most popular digital map companies in China. There are 65,848 POIs in the crawled data, including: company (43,225), school (3,946), residence (7,906), shopping (3,077), scenery (2,683), religion (318), health (4,693). Health POIs are subdivided into five categories: general hospital (380), specialty hospital (615), community hospital (1,520), animal hospital (318), and pharmacy (1,860).

## 3.2. Development period of bike sharing

Under the influence of various factors, the operation status of bike sharing is constantly changing. The monthly trip and user amounts of SBBS in Nanjing are shown in Figure 2. The four years are accordingly divided into four development periods: (1) SBBS expanding period, from March 2016 to February 2017; (2) DBS shock period, from March 2017 to August 2017; (3) stable period, from September 2017 to December 2019; (4) COVID-19 shock period, from January 2020 to February 2020.

In SBBS expanding period, user demand for SBBS travel grew rapidly, with a substantial increase in supply facilities. During that period, bike amount increased from 28,000 to 39,000 with an increase rate of 39%, and trip amount increased by an average of 122,800 per month. Increasing the supply of bikes and stations has a positive effect on promoting travel demand for bike sharing. In the DBS shock period, SBBS suffered strong competition from DBS. There was a significant decline in SBBS travel demand after the DBS promotion. In August 2017, Nanjing government banned allocating new DBS bikes because of bike over-supply, and then the SBBS operation has stabilized. In the stable period, SBBS survived the tide of DBS and still met extensive user demand. For example, SBBS attracted 323,600 users and generated 29.6 million trips in 2019. Besides, user demand is mainly affected by seasonal factors. User demand in spring and fall is the highest, and user demand in summer is lower. User demand in winter is the lowest because of the cold weather and the CNY holiday.

COVID-19 has a huge impact on bike sharing services. The amounts of trip and user have decreased significantly, and the decline was much greater than that in previous winters. The detailed analysis will be explained in the subsequent sections.

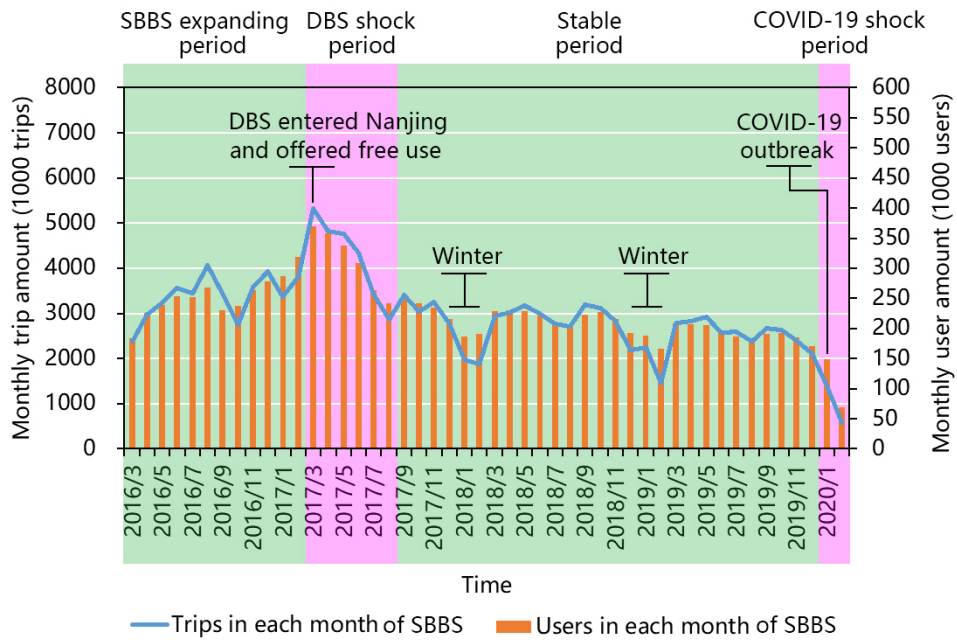

Figure 2. Development period of bike sharing in Nanjing.

### 3.3. Study period and spatial object

On January 20, 2020, Professor Nanshan Zhong, the leader of the high-level expert group of China's National Health Commission, affirmed the person-to-person transmission of SARS-CoV-2 (The Beijing News, 2020). The news immediately had a widespread impact in China. Jiangsu Province initiated a first-level response to public health emergencies, which lasted from January 25, 2020 to February 24, 2020. And 2020 CNY was January 25, 2020, which coincides with the COVID-19 outbreak. The impacts of COVID-19 and CNY need to be distinguished.

The detailed study period and corresponding trip amount are shown in Table 1. February 2020 is the most critical and most affected month, so the study period of COVID-19 impact is the post-period 2020/2/2-2020/2/15. The pre-period 2020/1/5-2020/1/18 will serve as the benchmark before the outbreak. In the pre-period, there is no COVID-19 case found in Nanjing; in the post-period, there are many COVID-19 cases found in Nanjing almost every day.

The post-period is the second and third weeks after the 2020 CNY, and the pre-period is the second and third weeks before the 2020 CNY. The data analysis of previous years' trips shows that travel demand would decrease before and after a week of the CNY, so the CNY impact can be ignored in the pre-period and post-period. Therefore, the pre-period and post-period are suitable for the before and after study of COVID-19 in Nanjing. Besides, the corresponding periods of 2017 are also analyzed to consider seasonal factors, because the weather conditions around 2017 CNY and 2019 CNY are similar.

Table 1. Trips of NYB in different periods.

| Service and year | Pre-period | Trips (1,000 trips) | Post-period | Trips (1,000 trips) |
|---|---|---|---|---|
| SBBS 2017 | 2017/1/8-2017/1/21 | 1950.2 | 2017/2/5-2017/2/18 | 2028.5 |
| SBBS 2020 | 2020/1/5-2020/1/18 | 735.1 | 2020/2/2-2020/2/15 | 213.1 |
| DBS 2020 | 2020/1/5-2020/1/18 | 2996.6 | 2020/2/2-2020/2/15 | 537.2 |

The spatial object of SBBS is the NPB station. The spatial object of DBS is the virtual station. Voronoi diagrams and buffer are combined to identify DBS virtual stations. Voronoi diagram of a given set of sites is the partition of a plane into regions close to each site. Firstly, the DBS service in Nanjing is partitioned into Voronoi regions centered by NPB stations. Then 250-m buffer of each NPB station and the corresponding Voronoi region are fused to get the intersection, which is the service area of each NPB station and corresponding virtual station. The service area of all stations is only 16% of Nanjing city area, but 70% of DBS trips are covered. It indicates that virtual station results are the suitable space objects of DBS analysis. The spatial distribution of virtual stations is shown in Figure 3(b).

## 4. Demand analysis: station perspective

### 4.1. Trip decline at station level

The difference-in-differences (DID) model is one of the most powerful and popular methods to evaluate the impact of extraordinary events and policy implementations. To estimate the influence of COVID-19 on SBBS trip, the traditional DID model and the semi-logarithmic DID model are both applied in the data analysis. The traditional DID model is represented in Equation (1), and the semi-logarithmic DID model is represented in Equation (2). The comparison of before-after periods and treatment-control groups are combined in the empirical methodology. CNY is the basis for judging the before or after period. 2017 observations are used as a control group, and 2020 observations are a treatment group.

$$Y_t = \varphi_t + \beta \cdot Year + \delta \cdot Post + \gamma \cdot Year \cdot Post + \varepsilon_t \tag{1}$$

$$\ln Y_t = \varphi_t + \beta \cdot Year + \delta \cdot Post + \gamma \cdot Year \cdot Post + \varepsilon_t \tag{2}$$

where the dependent variable $Y_t$ is the daily trip of Nanjing SBBS (in 1,000 trips), and the dependent variable $\ln Y_t$ is the logarithm of the daily trip. The vector $\varphi_t$ is a set of time-related dummy variables reflecting the time-varying trend, including the distance to CNY and the day of week. Parameter $\beta$ is a treatment group effect of different years. The dummy variable *Year* is defined as 1 for 2020 observations, and 0 for 2017 observations. Parameter $\delta$ is the post-period effect after CNY. The dummy variable *Post* is defined as 1 for post-period after CNY, and 0 for pre-period before CNY. Parameter $\gamma$ captures the impact of COVID-19 outbreak on SBBS trip.

The Semi-logarithmic DID model is found to be better than the traditional DID model. Because R-squared of semi-logarithmic DID model with 0.964 is large than that of traditional DID model with 0.931. In the estimated results of the semi-logarithmic DID model, the estimated values of $\beta$, $\delta$, and $\gamma$ are -1.079, -0.060, -1.289 respectively. The impact of different events on SBBS trips could be estimated by the values of these parameters. The SBBS trips in 2020 decreased by 66% compared to 2017, which can be interpreted as the DBS shock. The post-period effect after CNY would reduce SBBS trips by 6%. The most important finding is that the SBBS trips in Nanjing fell by 72% because of the COVID-19 impact. COVID-19 outbreak is the biggest challenge for SBBS during the four-year operation, followed by the DBS impact.

The semi-logarithmic DID model of *i*-th SBBS station is represented in Equation (3). Decrease rate of trip amount is the negative value of relative deviation between actual value during COVID-19 outbreak and expected value assuming no COVID-19. Due to the lack of 2017 DBS data, the formulas for calculating the decrease rate of SBBS and DBS are different. The expected trips of the SBBS station assuming no COVID-19 are represented in Equation (4). The expected trips of the DBS virtual station assuming no COVID-19 are the trips in the fortnight before 2020 CNY. The decrease rate of SBBS trip $d_i$ is represented in Equation (5), and the decrease rate of DBS trip $v_i$ is represented in Equation (6).

$$\ln Y_{i,t} = \alpha_i + \beta_i \cdot Year + \delta_i \cdot Post + \gamma_i \cdot Year \cdot Post + \varepsilon_{i,t} \tag{3}$$

$$\ln y_{i,2020post} = \alpha_i + \beta_i \cdot Year + \delta_i \cdot Post + \varepsilon_{i,t} \tag{4}$$

$$d_i = \frac{y_{i,2020post} - Y_{i,2020post}}{y_{i,2020post}} \tag{5}$$

$$v_i = \frac{X_{i,2020pre} - X_{i,2020post}}{X_{i,2020pre}} \tag{6}$$

where $\ln Y_{i,t}$ is the logarithm of SBBS trip in $i$-th station. Other variables and parameters in Equation (3) are similar to those in Equation (1), except $\alpha_i$ capturing the individual fixed effects at the station level. For the $i$-th station, $y_{i,2020post}$ is the expected amount of SBBS trips assuming no COVID-19, and $Y_{i,2020post}$ is the actual value of SBBS trips. $X_{i,2020pre}$ is the expected value of DBS trips assuming no COVID-19, and $X_{i,2020post}$ is the actual value of DBS trips.

The decrease rate of SBBS station trip during the COVID-19 outbreak is shown in Figure 3(a). SBBS trips at all stations have declined to varying degrees, and most stations have experienced a trip drop of around 70%. Xuanwu, Gulou, and Qinhuai districts are close to the city center and are the urban area of Nanjing. Qixia, Jianye, and Yuhua districts are suburban areas. Jianye and Qixia districts are the regions hit hardest by the pandemic. The station ratio with a high decrease rate in suburban areas is much larger than that in urban areas. Urban area is relatively less impacted by COVID-19, and suburban area is relatively more impacted. But there are some hot spots in the suburban area with a lower decrease rate.

Because of the COVID-19 impact, the DBS trip in Nanjing fell by 82%, which is larger than that of SBBS (72%). And 78% of DBS virtual stations' decrease rate is higher than the corresponding SBBS station. DBS is more severely affected by the pandemic than SBBS in most regions. As shown in Figure 3(b), DBS trips at all virtual stations have declined to varying degrees. DBS in the Jianye district is hit hardest by the pandemic among all districts. The DBS virtual stations with a high decreased rate coincide with the metro route, which shows that the integrated trip mode of bike-sharing-metro has been greatly affected by the pandemic. Because the metro is a confined space and has a higher transmission risk, the public is not prone to take a metro. The major role of bike sharing in urban transportation has changed from cooperating with public transport to independently providing mobility services.

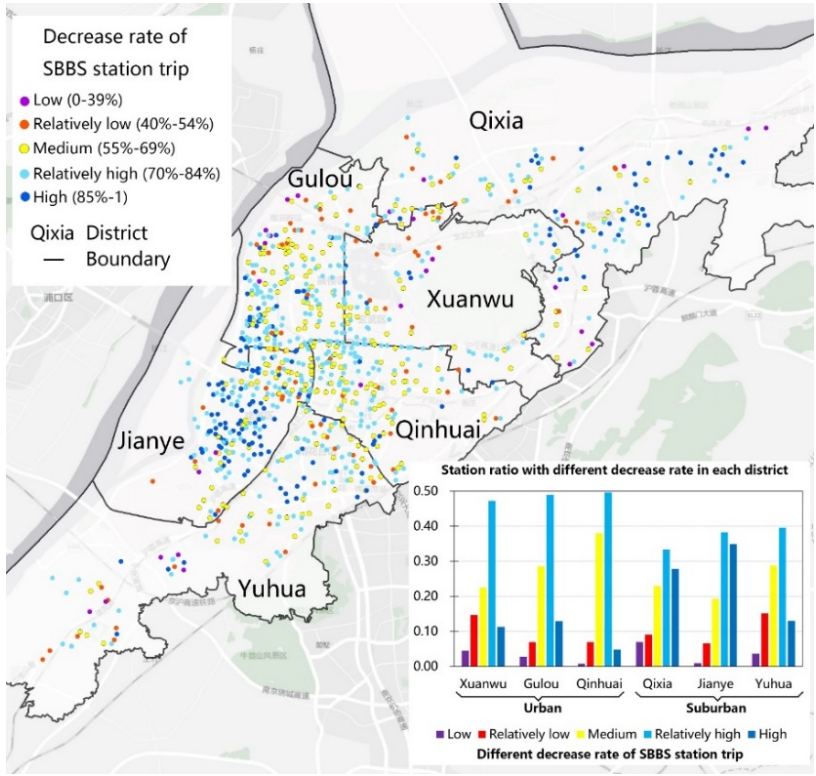

(a) The decrease rate of SBBS station trip.

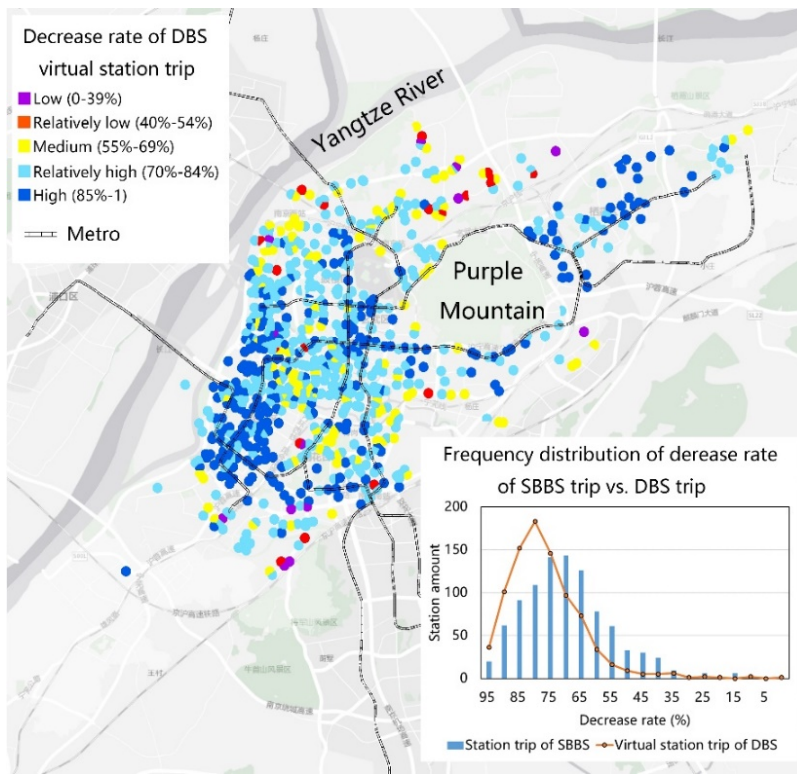

(b) The decrease rate of DBS virtual station trip

Figure 3. The decrease rate of bike sharing trips.

## 4.2. Land use and travel change

Land use could be represented by the distribution of POIs, which is closely related to travel purpose. The POI index is defined as the average amount of each type of POI near each trip, as shown in Equation (7). The POI index is proposed to show the relationship between travel demand and land use. The changes in POI index before and after the COVID-19 outbreak can reflect changes in the travel purpose of bike sharing users. For example, a decrease in the company POI index indicates that users are less likely to use bike sharing services for commuting.

$$I_k = \frac{\sum_{y=1}^{Y} p_k(O_y) + \sum_{y=1}^{Y} p_k(D_y)}{2Y} \tag{7}$$

where $I_k$ is the POI index of $k$-th type, $Y$ is all trips, $O_y$ and $D_y$ are the y-th trip's origin station and destination station, $p_k(S)$ is the amount of $k$-th type of POI at the service area of station $S$.

According to the POI index results in Table 2, the SBBS user's travel demand for commuting has decreased significantly, and the DBS user's travel demand for commuting and school has decreased significantly. SBBS's school trips increase and DBS's school trips decrease. Many school-related SBBS users may be school staff who still work during the outbreak, while the school-related DBS users are students and parents who need not go to school. The travel demand of SBBS and DBS for shopping and scenery has increased slightly. In both SBBS and DBS systems, the travel demand for health, religion, and residence has increased significantly. So areas near residential, religious, and health POI require more frequent disinfection by bike sharing companies. Besides, the residential POI index is about leaving or returning home, and its increase indicates a decrease in the diversification of travel purposes.

Healthcare service is very important during the pandemic. In both SBBS and DBS systems, the travel demand for pharmacy (buying medicines) has increased the most during the COVID-19. The increase of travel demand for community hospital is the second-highest for SBBS, and the increase of travel demand for specialized hospital is the second-highest for DBS. Travel change near the animal hospital is relatively insignificant. General hospital has a fever clinic and usually is crowded with many people, so higher transmission concerns have led to smaller increases in travel demand. Areas near hospitals or pharmacies should receive key guarantees for service supply and bike disinfection.

Table 2. POI index change of SBBS and DBS.
(a) POI index change of SBBS and DBS

| POI index | Company | School | Residence | Shopping | Scenery | Religion | Health |
|---|---|---|---|---|---|---|---|
| SBBS 2020 pre | 19.191 | 1.334 | 3.533 | 1.440 | 0.952 | 0.036 | 2.258 |
| SBBS 2020 post | 18.500 | 1.430 | 3.917 | 1.565 | 1.041 | 0.039 | 2.548 |
| Relative deviation of SBBS POI index | -3.6% | 7.2% | **10.9%** | 8.7% | 9.3% | **10.0%** | **12.8%** |
| DBS 2020 pre | 28.414 | 1.866 | 4.096 | 1.890 | 1.478 | 0.040 | 2.774 |
| DBS 2020 post | 26.072 | 1.670 | 4.521 | 1.993 | 1.498 | 0.043 | 3.092 |
| Relative deviation of DBS POI index | -8.2% | -10.5% | **10.4%** | 5.5% | 1.4% | **9.0%** | **11.5%** |

(b) POI index change of different health POIs.

| POI index | General hospital | Specialist hospital | Community hospital | Animal hospital | Pharmacy |
|---|---|---|---|---|---|
| SBBS 2020 pre | 0.283 | 0.486 | 0.507 | 0.162 | 0.821 |
| SBBS 2020 post | 0.316 | 0.542 | 0.576 | 0.169 | 0.945 |
| Relative deviation of SBBS POI index | 11.7% | 11.5% | **13.7%** | 4.5% | **15.2%** |
| DBS 2020 pre | 0.427 | 0.632 | 0.609 | 0.161 | 0.945 |
| DBS 2020 post | 0.473 | 0.714 | 0.665 | 0.165 | 1.075 |
| Relative deviation of DBS POI index | 10.8% | **13.1%** | 9.1% | 2.4% | **13.7%** |

## 5. Demand analysis: user perspective

### 5.1. Travel characteristics

The travel characteristic indexes of bike sharing are shown in Table 3. The number of SBBS users has dropped by 67%, and the travel frequency of remained SBBS users has decreased by 12%. The main reason is that the social distancing strategy encourages people to stay at home as much as possible in response to the pandemic. But the decrease in travel frequency does not mean that bike sharing has become less important. Before and after the COVID-19 outbreak, the daily trips of Nanjing Metro fell from 3,172,000 (pre-period) to 156,000 (post-period), with a drop of 95%. And bike sharing trips in Nanjing only fell by 72% for SBBS and 82% for DBS. Considering the higher decrease rate of other transport modes, bike sharing has played a more important role in emergency services of urban transportation.

The average travel distance of SBBS increased by 32%, and the average travel distance of DBS

increased by 16%. Public transport such as metro and bus is a confined space and has a higher transmission risk. As a response strategy, public transport in Nanjing has reduced departure frequency and shortened service time. Therefore, some long-distance travel demands are satisfied by bike sharing especially SBBS. This also confirms the previous finding that bike-sharing-metro mode declined because of the COVID-19 impact. A possible reason for this trip decline is that most bike-sharing-metro trips are for commuting purposes (Ma et al., 2018) but commuting is greatly impacted by the pandemic. Travel time increased accordingly, and there is little change in travel speed before and after the outbreak. DBS trips are faster than SBBS trips, because DBS users are younger than SBBS users. Almost every person wears PPE such as a mask during the pandemic, but wearing PPE did not significantly reduce the travel speed of bike sharing.

Network density is the ratio of the actual number of connections (origin-destination pairs) to all possible connections (Wise, 2014). Degree is the actual number of connections of each station (Saberi et al., 2018). Network density (value range is 0-1) and average degree can be used to show the network connectivity of user travel. The larger network density, the higher network connectivity. Network density of SBBS and DBS decreased by 45% and 43% respectively, reflecting that network connectivity of user travel experienced a great decline.

Table 3. The travel characteristic indexes of bike sharing before and after the COVID-19 outbreak.

| Characteristics | SBBS 2020 pre | SBBS 2020 post | DBS 2020 pre | DBS 2020 post |
|---|---|---|---|---|
| Daily trips (1,000 trips) | 52.3 | 15.1 ↓ | 214.0 | 38.3 ↓ |
| User amount (1,000 users) | 118.5 | 39.0 ↓ | - | - |
| Travel frequency (trips per week) | 3.09 | 2.72 ↓ | - | - |
| Travel distance (km) | 1.32 | 1.74 ↑ | 1.25 | 1.45 ↑ |
| Travel time (min) | 14.5 | 18.3 ↑ | 9.5 | 13.1 ↑ |
| Travel speed (km/h) | 7.40 | 7.54 ↑ | 8.57 | 8.12 ↓ |
| Average degree | 83.0 | 45.6 ↓ | 977.7 | 55.5 ↓ |
| Network density | 0.097 | 0.054 ↓ | 0.115 | 0.065 ↓ |

*\* Due to the lack of DBS user information, user amount and travel frequency of DBS are blank.*

The temporal characteristics of bike sharing trips are shown in Figure 4. There are significant morning peak and evening peak for bicycle sharing before the pandemic. The morning peak is 7 am to 9 am and the evening peak is 5 pm to 7 pm. These peaks became less obvious during the pandemic, reflecting a huge decrease in commuting travel. The peak phenomenon of DBS is more significant than that of SBBS before the outbreak, indicating that commuting travel accounts for a larger proportion of DBS trips. After the outbreak, the temporal characteristics of DBS and SBBS become similar. So DBS is more affected by COVID-19 than SBBS. No matter before or after the outbreak, the travel demand on weekdays is higher than the travel demand on weekends. It should be noted that the vertical axis of the four subgraphs in Fig. 4 are different, and the colors of each sub-graph cannot be compared with the others. For example, although the color of Figure 4 (a) is lighter than that of Figure 4 (b), the trip amount of SBBS 2020 pre is much higher than that of SBBS 2020 post.

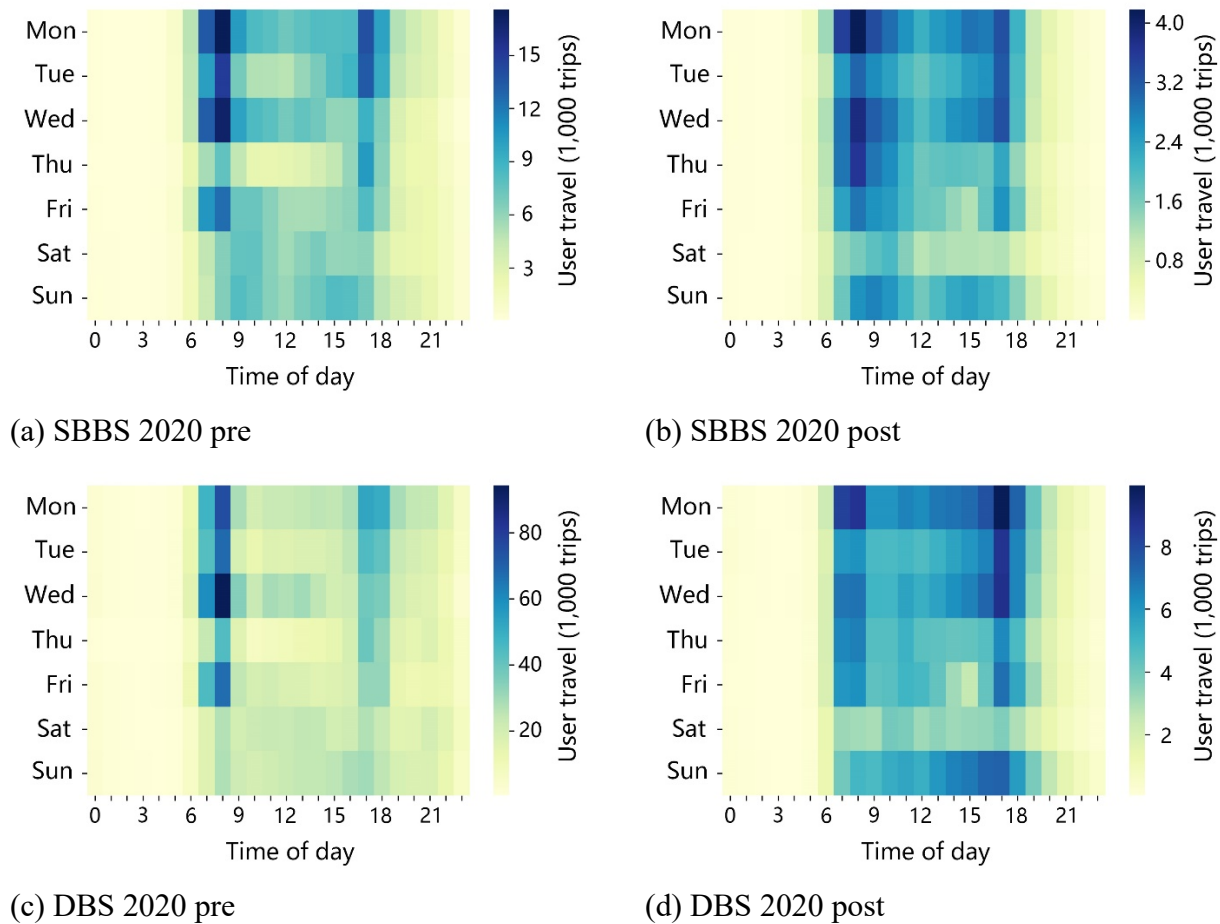

(a) SBBS 2020 pre

(b) SBBS 2020 post

(c) DBS 2020 pre

(d) DBS 2020 post

Figure 4. Temporal patterns of bike sharing trips.

## 5.2. Gender and age group

Due to the lack of DBS user information, the analysis of gender and age focus on SBBS user travel. The travel speed of males (7.6 km/h) and females (7.2 km/h) has not been changed by the pandemic. The bike sharing users are divided into five age groups: teenage (13-18 years old), youth adult (19-28 years old), adult (29-39 years old), middle age (40-59 years old), the elderly (≥60 years old). The female and male trips of different ages are shown in Figure 5. Before the COVID-19 outbreak, the trips of females and males are roughly equal, except that the elderly trips of females are less than that of males. But the middle age trips of females are also less than that of male after the outbreak. The female trip proportion in all trips fell from 47% to 43%, which shows that women are more affected by the pandemic than men. It also means that men are more dependent on bike sharing than women.

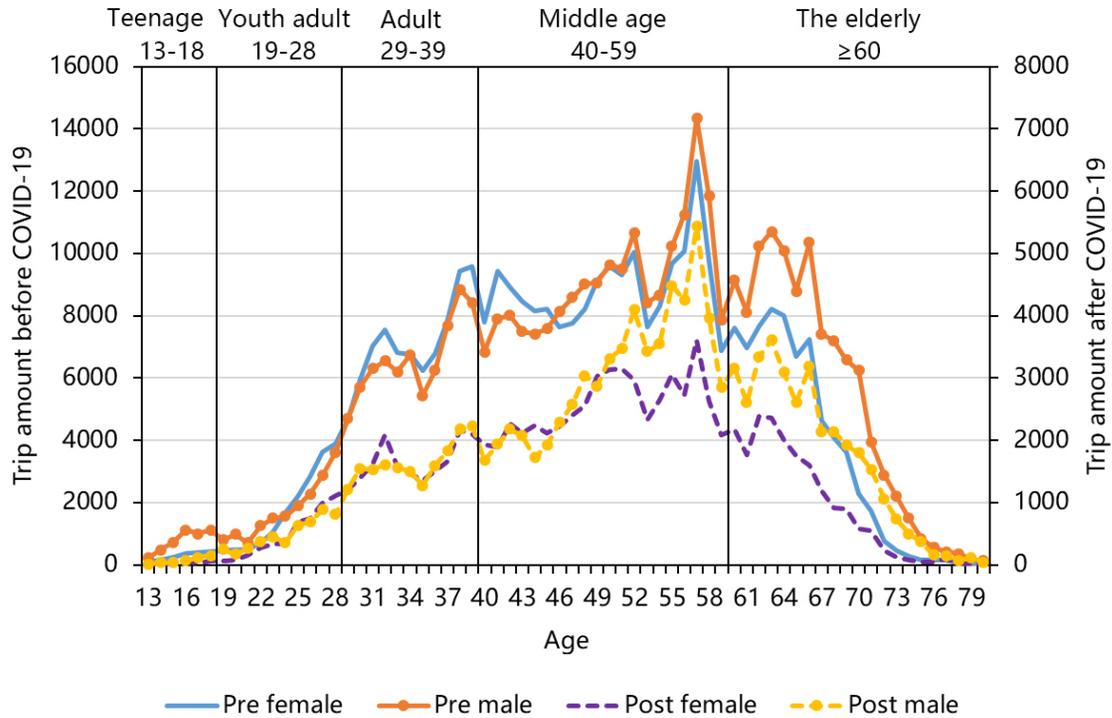

Figure 5. The SBBS trip amount in relation to gender and age.

The SBBS travel changes for different age groups are shown in Table 4. The proportion of teenage users is the lowest, and their travel demands are also the hardest hit by the COVID-19. The user proportion of youth adult and adult have slightly decreased, and the decline in their travel demand is relatively small. The proportion of middle-aged and elderly users has increased, and their trip amount has been least affected by the pandemic. The travel frequency of middle age and the elderly are also highest during the pandemic. These findings show that middle-aged and elderly people are the key groups of SBBS users and have a strong dependence on bike sharing services. Middle-aged and elderly users should get more attention from bike sharing companies, especially elderly users who are more susceptible and vulnerable to the coronavirus.

Table 4. The SBBS travel changes of different age groups before and after COVID-19.

| Age group | User proportion | | Travel frequency (weekly trips) | | Decrease rate of trip amount |
|---|---|---|---|---|---|
| | 2020 pre | 2020 post | 2020 pre | 2020 post | |
| Teenage | 1.1% | 0.4% ↓ | 2.55 | 1.78 | 90.6% |
| Youth adult | 4.9% | 4.7% ↓ | 3.01 | 2.63 | 72.2% |
| Adult | 22.4% | 18.8% ↓ | 2.86 | 2.47 | 76.1% |
| Middle age | **49.8%** | **51.4% ↑** | 3.05 | **2.80** | **68.9%** |
| The elderly | 21.9% | **24.6% ↑** | **3.46** | **2.79** | **70.2%** |
| All | - | - | 3.09 | 2.72 | 72.0% |

The POI index is used to analyze the relationship between land use and the SBBS trips of various age groups. The relative deviation of the POI index is shown in Table 5. The importance of shopping and health especially pharmacy has increased significantly in trips of all age groups, while the importance of commuting travel has decreased significantly in all age groups. Teenage users have significantly reduced travel for school, scenery, and religion, but other age groups have increased travel for these purposes. During the pandemic, teenagers avoid going to general hospitals and specialized hospitals, which may result from parents' safety concerns. The travel purposes of youth adult and adult users are greatly affected by COVID-19. But the travel purposes of middle age and elderly users are less affected by the pandemic.

TABLE 5. The relative deviation of SBBS POI index of different age groups.

| POI | Teenage | Youth adult | Adult | Middle age | The elderly | All |
|---|---|---|---|---|---|---|
| Company | -17.1% | -4.6% | -6.9% | -3.1% | -2.6% | -3.6% |
| School | -10.5% | 9.5% | 12.8% | 6.4% | 3.4% | 7.2% |
| Residence | 10.2% | **15.7%** | **19.9%** | 8.5% | 7.6% | 10.9% |
| Shopping | 13.5% | 13.5% | 12.2% | 8.1% | 5.7% | 8.7% |
| Scenery | -16.7% | **21.6%** | **21.0%** | 6.5% | 4.4% | 9.3% |
| Religion | -4.7% | **42.9%** | **15.7%** | 6.4% | 7.2% | 10.0% |
| Health | 7.4% | **17.1%** | **18.8%** | 10.9% | 10.8% | 12.8% |
| General hospital | -23.4% | **15.4%** | **21.5%** | 6.3% | 14.4% | 11.7% |
| Specialist hospital | -17.9% | **17.9%** | **21.8%** | 8.3% | 9.1% | 11.5% |
| Community hospital | 2.6% | **21.1%** | **18.7%** | 12.2% | 11.5% | 13.7% |
| Animal hospital | **27.7%** | 8.5% | 5.6% | 2.5% | 6.4% | 4.5% |
| Pharmacy | **37.5%** | **16.2%** | **19.0%** | 15.0% | 11.1% | **15.2%** |

# 6. Risk analysis: bike perspective

## 6.1. Transmission risk and bike usage interval

Unlike the confined space of metro or car, bike sharing service is provided in the open space. So the main concern of virus transmission risk is the close contact by touching the bike surface. Bike usage interval is defined as the time interval between former user travel and latter user travel of the same bike, which also is the time interval of user contact. According to this definition, the amount of bike usage interval is equal to the amount of user contact. Bike usage interval can be used to characterize the user contact and the transmission risk.

During the COVID-19, SBBS bike turnover rate decreased from 1.7 trips per day to 0.8 trips per day, and DBS bike turnover rate decreased from 1.0 trips per day to 0.4 trips per day. The decrease of bike turnover rate directly leads to the duration increase and amount decrease of bike usage interval. After the COVID-19 outbreak, the average value of SBBS bike usage interval increased from 8.7 hours to 14.4 hours, and the average value of DBS bike usage interval increased from 12.5 hours to 23.7 hours. The daily amount of SBBS user contact decreased from 46,600 times to 11,700 times, and the daily amount of DBS user contact decreased from 199,400 times to 31,200 times. The user contact of all companies in Nanjing is 42,900 times per day. Nanjing has four bike sharing companies, NPB, Mobike, Hellobike, and Didibike. The bike disinfection of one company is about 20,000 times per day, so the bike infection of all four companies could exceed the amount of user contact.

The frequency distribution of bike usage interval is shown in Figure 6. After the COVID-19 outbreak, the bike usage interval of SBBS and DBS both are greatly reduced. Before the outbreak, there were two peaks every 24 hours for the frequency of bike usage interval. The first peak is the 8th hour (from morning peak 9 am to evening peak 5 pm), and the second peak is the 24th hour (the next day's travel). After the outbreak, the bike usage interval has only one peak every 24 hours, representing the travel of next day. This finding is in mutual corroboration with the temporal visualization in Figure 4. Bike usage interval is determined by travel demand, and travel demand drop and weaker peak phenomenon are the reasons of the amount decrease and duration increase of bike usage interval. The possibility of user contact was further reduced, so trip decline indeed reduces the risk of virus transmission through bike sharing.

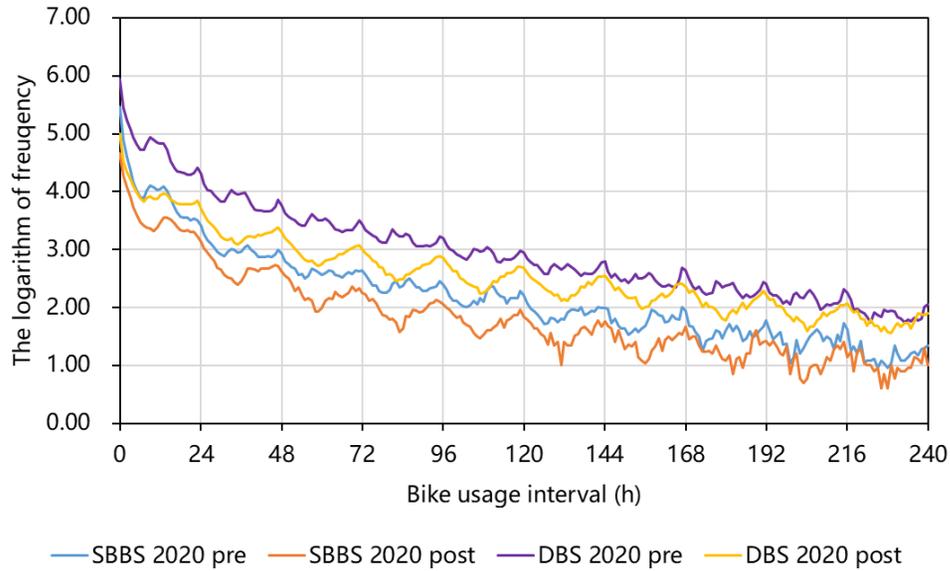

Figure 6. The frequency distribution of bike usage interval of SBBS and DBS.

When riding a bike, users will touch the handlebars and seats, which are made of stainless steel and plastic. Considering that the virus's half-life is about 6 hours on these materials (van Doremalen et al., 2020), we tentatively choose 6 hours as the safety threshold for bike usage interval. If a bike usage interval is less than the safety threshold, a user close contact occurs. After the COVID-19 outbreak, the daily amount of SBBS user close contacts decreased from 33,400 times to 6,800 times, with a decrease rate of 80%. And the daily amount of DBS user close contacts decreased from 115,400 times to 14,000 times, with a decrease rate of 88%. The decrease rate of user close contact is higher than that of bike sharing trips (SBBS: 80% vs. 72%, DBS: 88% vs. 82%). The spatial distribution of user close contact is shown in Figure 7. There are more user close contacts in central urban areas than that in suburban areas, which means a higher risk of virus transmission. The high-risk places should be given priority in bike disinfection.

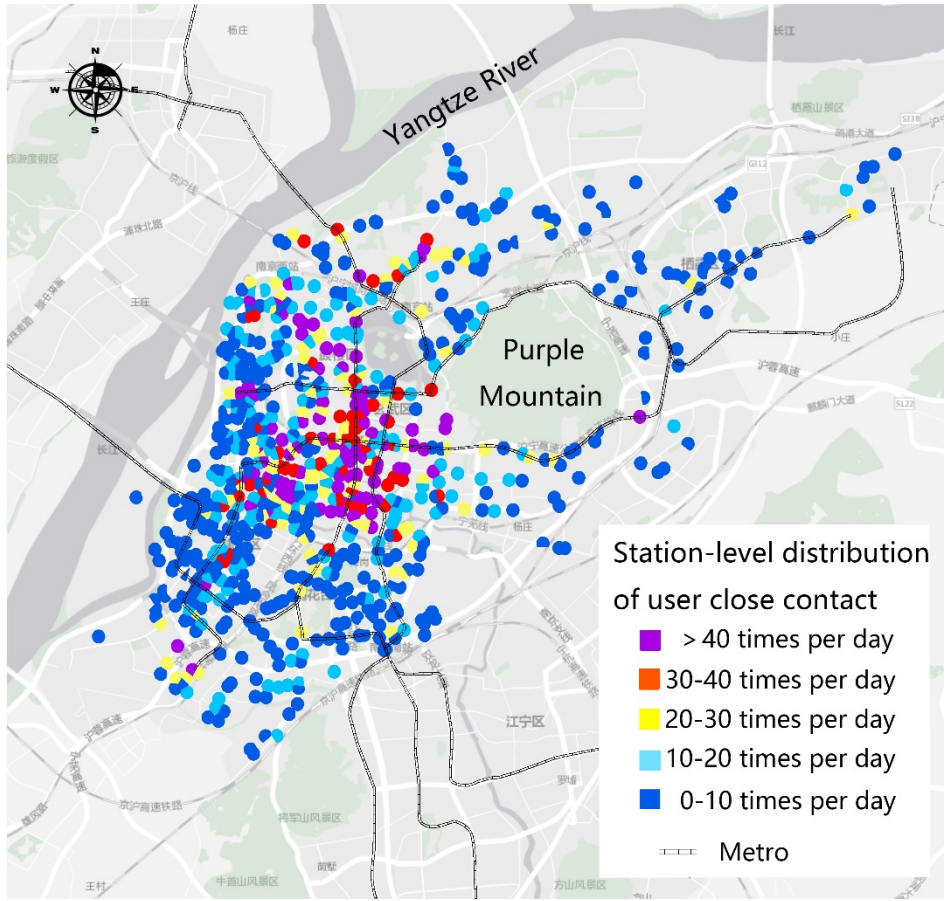

Figure 7. The station-level distribution of user close contact during the COVID-19.

## 6.2. Service supply and user distancing

Service supply of bike sharing consists of bike supply and operational resources. The statistic indexes of service supply and transmission risk in bike sharing are shown in Table 6. Bike supply is more sufficient as the travel demand declines during the COVID-19 pandemic. Idle bike ratio is the ratio of bikes that have not been used in a certain period to the all bike supply. The higher the idle bike ratio, the greater bike over-supply. The idle bike ratios of SBBS and DBS both are 42% before the outbreak, that is, only 58% of the bikes are used in these two weeks. After the outbreak, the idle bike ratio becomes even higher and bike over-supply becomes more serious. These idle bikes should be activated to meet travel demand and used to avoid user close contact.

Bike repositioning is the most important part of operational resource supply. Bike repositioning is to solve the imbalance between travel demand and bike supply, especially in the morning and evening peaks. The amount of bike repositioning experiences a huge reduction after the COVID-19 outbreak, because travel demand greatly decreases and the peak phenomenon becomes insignificant. As shown in Table 6, SBBS bikes are daily repositioned 6,300 times before the pandemic and 1,400 times after the outbreak, with a decrease of 78%. DBS bikes are daily repositioned 72,400 times before the pandemic and 10,900 times after the outbreak, with a decrease of 85%. The decrease rate of bike repositioning amount is large than that of bike sharing trip. During the pandemic, the company's repositioning objectives should shift from increasing bike efficiency to reducing transmission risk. The decline of repositioning amount results in a relative surplus of operational resources, and more operational resources could be devoted to bike disinfection.

Besides, we propose a new concept of *user distancing* to help avoid transmission risk. User distancing is an application of social distancing strategy and user information-sharing in shared micro-mobility. User distancing in bike sharing is to ensure that bike usage interval should remain above the safety threshold and user close contact should be avoided. The potential user of each bike is informed of the travel end time of the previous user and decides her/his travel behavior accordingly. For example, if the parking duration of a bike to be used is shorter than the safety threshold, the user will be informed that the bike is unavailable for travel and encouraged to ride another idle bike or a bike with enough parking duration. If bikes are parked in the high-demand area and require efficient turnover, the operation staff should be arranged to patrol the area and ensure timely disinfection. The relative surplus of bike supply and operational resources promotes the feasibility of user distancing strategy. The user distancing strategy can reduce the bike disinfection workload, reduce operational cost and increase the turnover efficiency of idle bikes.

Table 6. The resource supply and transmission risk in bike sharing.

| Sample | Bike supply (1,000 bikes) | Idle bike ratio (%) | Bike repositioning (1,000 times daily) | User close contact (1,000 times daily) |
|---|---|---|---|---|
| SBBS 2020 pre | 53.2 | 41.5 | 6.3 | 33.4 |
| SBBS 2020 post | | 64.9 ↑ | 1.4 ↓ | 6.8 ↓ |
| DBS 2020 pre | 350.6 | 41.6 | 72.4 | 115.4 |
| DBS 2020 post | | 71.9 ↑ | 10.9 ↓ | 14.0 ↓ |

## 7. Conclusion and discussion

The COVID-19 pandemic has had a huge impact on the globe, and transportation services in many cities have undergone significant changes. Bike sharing, as a kind of shared micro-mobility, played an important role in urban transportation during the pandemic. Many cities insist on providing bike sharing services, while some other cities have decided to suspend this service. Travel demand and transmission risk are the main concerns and the key basis for urban decision-making. Based on the trip data of SBBS and DBS in Nanjing, this study explored the usage pattern of the bike sharing system during the pandemic from three perspectives: station, user, and bike. The station and user perspectives are used to analyze travel demands, and the bike perspective is used to analyze virus transmission risk. This paper provides a valuable case study of providing SMM service during the COVID-19 pandemic.

From the station perspective, bike sharing has been greatly impacted by COVID-19, and travel demand for each (virtual) station has decreased variously. The decrease rate of SBBS trips (72%) is lower than that of DBS trips (82%), meaning that SBBS is less affected by the outbreak. The decrease rate in urban areas is lower than that in suburban areas, but some hot spots in suburban areas are less affected by the pandemic. Besides, bike sharing trips near the metro have significantly decreased, and travel distance has greatly increased after the outbreak. During the pandemic, bike sharing provides more independent and long-distance transportation services, rather than integrated services with public transport such as bike-sharing-metro mode. Besides, bike sharing trips for commuting have greatly decreased, but trips for health and religion have increased significantly. Health and religion POIs are the key areas of bike sharing services during the COVID-19, and the adjacent bike sharing services need to be guaranteed with priority.

From the user perspective, bike sharing has become more important during COVID-19, and middle-aged and elderly people are more dependent on this service. Bike sharing is far less affected by the pandemic than metro, and it replaces some of the long-distance travel. At the same time, the network connectivity and peak phenomenon of bike sharing have both weakened. Women are more impacted than men, especially middle-aged females. The teenagers' trips have decreased significantly, and their school trips have been transferred to other purposes. The travel purposes of youth adult and adult have changed greatly, with a marked decrease in commuting travel and an important increase in tourism, religion, and health travel. Middle-aged and elderly people are key user groups of SBBS, and their trip decline and travel purpose change are the smallest among all age groups.

From bike perspective, the decline of travel demand reduces user contact, and adequate bike disinfection can avoid the risk of virus transmission. Bike usage interval and user close contact are used to evaluate the transmission risk. Less travel demand and weaker peak phenomenon contribute to the amount decrease and duration increase of bike usage interval. The decline of user close contact in SBBS and DBS is 80% and 88%, which are both greater than the trip decline of SBBS and DBS. User close contact in the central urban area is higher than the suburban area, so the city center is the key area for bike disinfection. Under the influence of COVID-19, the service supply such as bike supply and operating resource is relatively surplus. Bike sharing companies can encourage users to ride idle vehicles to reduce user contact, and transfer some operating resources to bike disinfection.

During the pandemic, SMM such as bike sharing can both provide safe service and avoid transmission risk, with appropriate control strategies. These management strategies include social distancing, wearing PPE, disinfecting vehicles, and washing hands in time. Chinese cities such as Nanjing and Wuhan insisted on providing bike sharing services during COVID-19. While the pandemic in these cities has been effectively controlled and there is no report about bike sharing spreading virus. These successful practices prove that bike sharing can simultaneously meet travel demand and avoid safety risks. The answer to the question in the title is that cities with good management are encouraged to maintain SMM services such as bike sharing during the outbreak.

COVID-19 is unlikely to end in the short term, and the pandemic control strategies in the long term needs to be considered. Strict strategies such as city shutdown adopted earlier are not sustainable in the long term. Social life will continue to recover, and travel restrictions will gradually be relaxed. With increased travel, people contact will become more intensive. Therefore, how to effectively control the virus transmission risk in the long term is a huge challenge. Reliable, low-cost, and flexible pandemic control strategies in urban transportation should be explored. For example, we have proposed that companies can adopt the user distancing strategy based on activating idle vehicles. Besides, users should be encouraged to wear PPE and disinfect bikes before riding, and maintain these habits for a longer time.

## Acknowledgements

This research is sponsored by the National Key Research and Development Program of China (2018YFB1601300) and the Graduate Research and Innovation Projects of Jiangsu Province (KYCX20_0132). The authors appreciate Nanjing Public Bicycle Co., Ltd. and Nanjing Transportation Bureau for providing the data used in this study.